\input{aipcheck}

\documentclass[final            
  ]
  {aipproc}

\layoutstyle{6x9}

\begin{document}

\title{Radial Velocity search for substellar companions to sdB stars}


\author{Simon O'Toole}{
  address={Australian Astronomical Observatory, PO Box 296, Epping 1710, Australia}
}

\author{Uli Heber}{
  address={Dr Remeis-Sternwarte, Universit\"at Erlangen-N\"urnberg, Sternwartstrasse 7, Bamberg, D-96049, Germany}
}

\author{Stephan Geier}{
  address={Dr Remeis-Sternwarte, Universit\"at Erlangen-N\"urnberg, Sternwartstrasse 7, Bamberg, D-96049, Germany}
}

\author{Lew Classen}{
  address={Dr Remeis-Sternwarte, Universit\"at Erlangen-N\"urnberg, Sternwartstrasse 7, Bamberg, D-96049, Germany}
}

\author{Orsola De Marco}{
  address={Dept of Physics and Astronomy, Macquarie University}
}

\begin{abstract}
After the discovery of a substellar companion to the hot subdwarf
HD\,149382, we have started a radial velocity search for similar
objects around other bright sdB stars using the Anglo-Australian
Telescope. Our aim is to test the hypothesis that close substellar
companions can significantly affect the post-main sequence evolution
of solar-type stars. It has previously been proposed that binary
interactions in this scenario could lead to the formation of hot
subdwarfs. The detection of such objects will provide strong evidence
that Jupiter-mass planets can survive the interaction with a
solar-type star as it evolves up the Red Giant Branch. We present
the first results of our search here.
\end{abstract}

\maketitle


\section{Introduction and motivation}

Most investigations into the so-called ``Hot Jupiters'' and other
exoplanets close to their parent stars have focussed on the formation
and migration of these objects to their present-day location. The
planets' subsequent evolution -- and especially their effect on the
evolution of the stars they orbit -- has received less attention. For
the former, the proximity of the exoplanet to its star leads to
measurable mass loss through evaporation (e.g. Vidal-Madjar et
al.\ 2003). By considering the energy of the system, Lecavelier des
Etangs (2007) found that despite this mass loss, all of the known
exoplanets will survive at least 5 billion years. On these time-scales
the evolution of the host stars begins to become important.

In a study examining the influence of planets on post-main sequence
evolution, Soker (1998) found that substellar companions in orbits of
up to 5\,AU interact with the evolving star as it expands during the
red giant phase. Mass loss on the red giant branch is enhanced as the
companion(s) deposit angular momentum and energy in to the stellar
envelope and this leads to a bluer horizontal branch (HB) star than
might otherwise be expected. Soker used this model to explain the
observed morphologies of the HB in galactic globular clusters, and
predicted that massive planets or brown dwarfs should orbit stars at
the extreme blue end of the HB with orbital periods of $\sim$10
days. In a later study, Livio \& Soker (2002) found that at least
3.5\% of evolved solar-type stars will be ``significantly affected by
the presence of planetary companions''. This number increases to more
than 9\% for stars with metallicities above the solar value. It is now
well established that metal-rich stars are more likely to harbour
planetary companions (e.g. Fischer \& Valenti 2005).  An analysis of
the group properties of exoplanets of Marcy et al.\ (2008) found that
$\sim$4\% of solar-type stars have planets with orbits of
$<$2.5\,AU. Most recently, Bowler et al.\ (2010) found that
26$^{+9}_{-8}$\% of evolved A-type stars (1.5$\le M_*/M_\odot\le$2.0)
host Jupiter-mass planets within 3\,AU. It is clear then, that
\emph{there should be a population of very blue HB stars with
  substellar companions.} 

\section{The hot subdwarfs}

The very blue, or extreme, HB stars are the hot subdwarf B (sdB)
stars. These objects, like their more normal HB counterparts, are core
helium-burning stars, except with hydrogen envelopes too thin to
sustain nuclear burning. Their masses are typically
$\sim$0.5\,M$_\odot$. After the consumption of helium in their cores,
they evolve directly into white dwarfs, avoiding a second red-giant
phase. Most formation scenarios for sdB stars have focussed on close binary
interaction with a main sequence -- not substellar -- companion or the
merger of two He-core white dwarfs (e.g. Han et al.\ 2003). A large fraction
of sdB stars are predicted to be in close binaries with a main
sequence star or white dwarf companion.

Several radial velocity studies have found that this is the case: many
sdB stars reside in close binaries with periods as short as 0.07 days,
and with either an M-type main sequence star or an invisible white
dwarf companion (e.g. Maxted et al.\ 2001; Heber et al.\ 2004; O'Toole
et al.\ 2004; Edelmann et al.\ 2005). Other studies have used 2MASS
photometry to estimate the fraction of sdBs with main sequence stars
(Stark \& Wade 2004; Reed \& Stiening 2004), although they are limited
by the flux of the sdB to stars earlier than $\sim$M2. Binary fraction
estimates are in the 40-70\% range, with selection effects difficult
to determine. This still leaves at least 30\% of all sdBs as
apparently single stars. The Han et al.\ (2003) formation models
suggest that these stars are the product of a merger between two
helium-core white dwarfs. It is not clear, however, whether there are
enough of these double-degenerate systems that are close enough to
merge within a Hubble time. Perhaps instead of mergers,
the majority of single sdB stars are the product of common
envelope evolution with a \emph{substellar} companion. 

\section{The search for substellar companions to sdB stars}

The discovery of HD\,149382b with mass 6-23\,M$_{\mathrm{Jup}}$ by
Geier et al.\ (2009) has clarified the situation somewhat, and forced a
re-examination of the Soker (1998) and Livio \& Soker (2002)
models. The detected Doppler velocity variations of the sdB star are
sufficiently low ($K=2.3\,{\rm km\,s^{-1}}$, see Figure 1) that
previous surveys for RV variability -- whose limits are typically
2-3\,km\,$^{-1}$ -- would not have seen them. Furthermore, HD 149382
is the brightest known sdB, where very high quality data is easily
accessible. We note however, that this results is the subject of debate; see Jacobs et al. (these proceedings). The detection of more substellar companions in short-period
orbits around other sdB stars will strengthen the case that these
objects \emph{can} cause common envelope ejection.

\subsection{Using UCLES + CYCLOPS on the AAT}

We have been granted 10 nights in total with the Anglo-Australian Telescope (AAT) to carry out time-resolved high-resolution spectroscopy with UCLES/CYCLOPS of a sample of bright sdBs. The CYCLOPS fibre-feed to UCLES provides higher resolution (R=70000) with no loss in signal when compared to the standard mode (R$\approx$45000). One of the key features of the new system is the $\sim$2.1 arcsecond lenslet array feeding the fibres; this makes the spectrograph more immune to the sometimes poor seeing at Siding Spring Observatory. Overall a gain in throughput is expected once the system is fully implemented. Our goal is to search for Doppler velocity variations with semi-amplitudes of 1-2\,km\,s$^{-1}$ on timescales of days. This will allow us to detect companions with masses as low as $\sim$4M$_{\mathrm{Jup}}$. 

\subsection{Target Selection}

Previous observations of the bright sdB HD\,205805 with ESO-2.2m/FEROS found a shift of 2.5$\pm$0.5 km\,s$^{-1}$, larger than the measurement uncertainties. This object, along with HD\,149382, was one of our highest priority targets with UCLES/CYCLOPS. No intensity variations have been detected for this star (Chris Koen, priv. comm.) as might be expected for stellar pulsation, suggesting that the Doppler velocity variability is more likely due to a companion. We have taken high cadence observations of the star, which will allow us to accurately measure its orbital period. The other stars in our sample are well studied bright subdwarfs where no companion has been detected up to now, either with Doppler velocities or infrared colours. Should the ejection of the common envelope be caused by close substellar companions for apparently single sdBs, most of these stars are predicted to show Doppler velocity variations with low semi-amplitudes. The fact that the brightest known sdB shows variability is a strong hint in this direction. 

\begin{figure}{t}
  \includegraphics[angle=270,width=\textwidth]{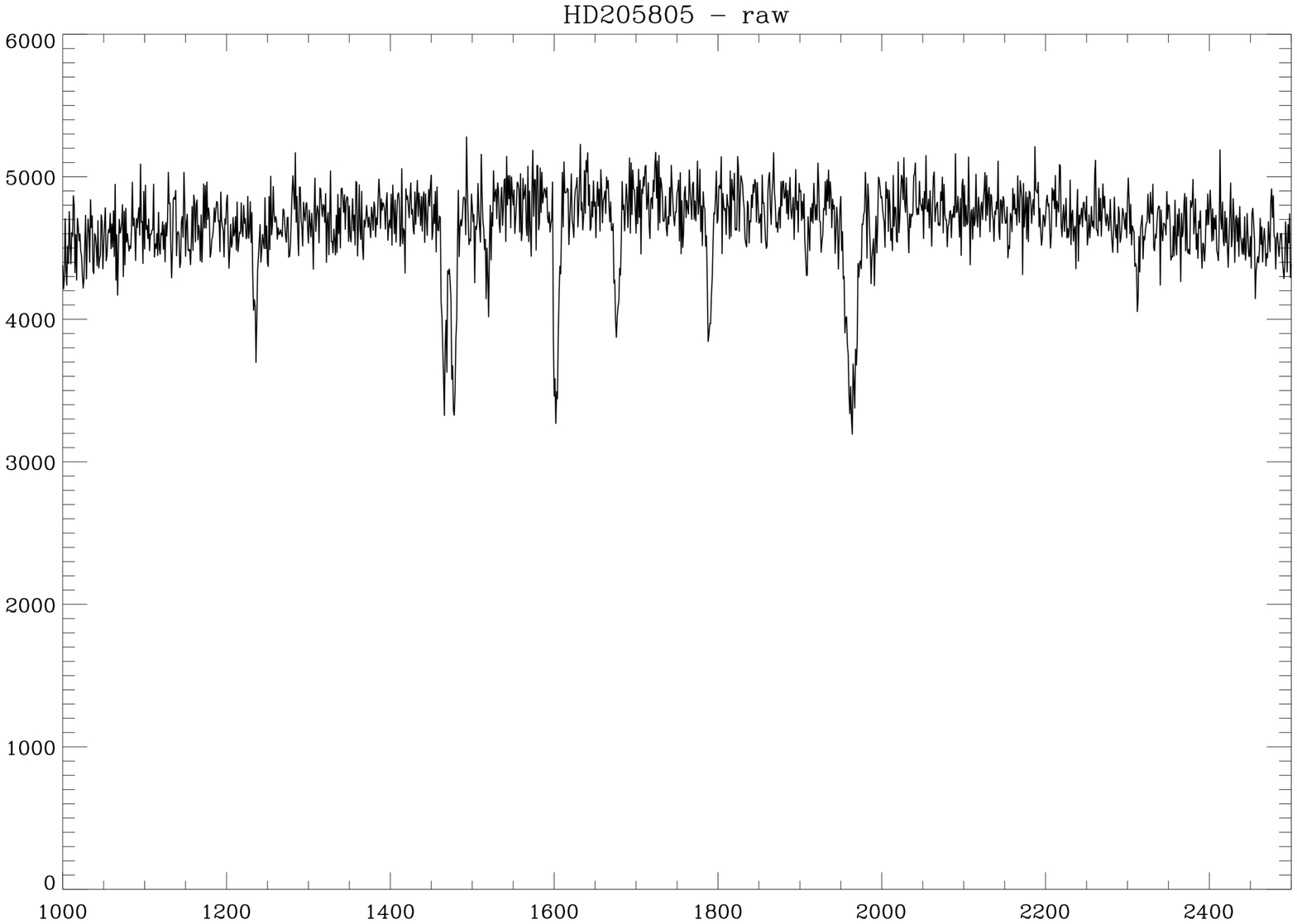}
  \caption{An extracted non-wavelength calibrated CYCLOPS spectrum of HD\,205805, showing pixels and counts on the x- and y-axes, respectively.}
  \label{fig:205805}
\end{figure}

\section{Early results... or what you will}

Our first four night allocation was very successful, despite a few small teething problems with the newly commissioned instrument (note that CYCLOPS was offered in shared risk mode for the first few semesters). We obtained over 50 spectra of HD\,149382, and over 30 of HD\,205805. An example of an extracted (but not yet wavelength calibrated) spectrum of the latter star is shown in Figure \ref{fig:205805}. We also have time-series spectra of another 3-4 objects. Once we have determined a wavelength solution for the spectra we will be in an excellent position to address some of the questions raised here.





\begin{theacknowledgments}
We would like to thank Chris Tinney for his help using CYCLOPS, Paul Butler for reducing the raw CYCLOPS data, and the staff at the AAT for help debugging the system.
\end{theacknowledgments}



\bibliographystyle{aipprocl} 


\IfFileExists{\jobname.bbl}{}
 {\typeout{}
  \typeout{******************************************}
  \typeout{** Please run "bibtex \jobname" to optain}
  \typeout{** the bibliography and then re-run LaTeX}
  \typeout{** twice to fix the references!}
  \typeout{******************************************}
  \typeout{}
 }

\end{document}